# Transmission of O-band wavelength-division-multiplexed heralded photons over a noise-corrupted optical fiber channel

Mao Tong Liu[1] and Han Chuen Lim[1,2,*]

[1]*School of Electrical and Electronic Engineering, Nanyang Technological University, 50, Nanyang Avenue, 639798, Singapore*
[2]*Emerging Systems Division, DSO National Laboratories, 20, Science Park Drive, 118230, Singapore*
[*]*hanchuen@dso.org.sg*

Abstract

We transmitted O-band heralded photons over 10 km of optical fiber in a proof-of-concept experiment demonstrating the feasibility of using heralded photons to improve the noise tolerance of quantum key distribution. In our experiment, the optical fiber channel was corrupted by noise photons to the extent that if we had used an attenuated laser as the photon source, a photon signal-to-noise ratio of $< 4.0$ at the receiver, corresponding to a quantum bit-error rate of $> 10.0\%$, would have prevented the effective generation of secure keys. Using a photon heralding scheme, the photon signal-to-noise ratio in our experiment was shown to be $> 7.8$. This corresponds to a quantum bit-error rate of $< 5.7\%$, which is good enough for distilling secure keys. In addition, we showed that it is possible to incorporate wavelength-division-multiplexing into the photon heralding scheme to improve overall key rate. We discussed and clarified the prospects and limitations of the photon heralding scheme for noise-tolerant quantum key distribution.

## 1. Introduction

Quantum key distribution (QKD) is an emerging technology that has attracted considerable attention in recent years. Based on physics principles rather than computational assumptions, QKD has the potential to revolutionize the way we secure our communications. In the past decade, there have been many demonstrations of QKD systems in both lab and field environments. Most QKD systems today, including commercial off-the-shelf systems, are implemented over low-noise dedicated *dark fibers* that do not carry Internet data traffic [1–3]. Although desirable from a system implementation perspective, the use of dedicated fibers for QKD can be costly. A few groups have already demonstrated the feasibility of QKD over *bright fibers* through wavelength-multiplexing both QKD and Internet data channels within the low-loss *C-band* (1530–1565 nm) [4–7]. To prevent four-wave mixing (FWM) crosstalk and spontaneous Raman scattering (SRS) noise photons from corrupting the QKD channel, it is important to carefully select the Internet data channel wavelengths and limit their launched optical power. Moreover, tight temporal gating of photon detectors and narrowband spectral filtering at the receiver are shown to be needed for obtaining high key rates [7].

For intra-city secure key distribution between trusted network nodes that are separated by distances shorter than 40 km [5], one can consider using the O-band (1260–1360 nm) instead for QKD [8–13]. This wavelength choice is attractive because the QKD channel and C-band Internet data channels are wavelength separated by $> 170$ nm, and so the noise condition at O-band becomes less severe [8, 9]. It would be highly desirable if we could make QKD and Internet services coexist on the same fiber without having to impose any restrictions on the Internet services. This shall allow non-disruptive introduction of QKD into existing fiber-optic networks and accelerate wide-spread deployment of QKD systems.

With absolutely no restrictions imposed, a QKD service provider has limited control over the noise condition experienced by the QKD channel. It is therefore possible that noise photons originating from SRS of Internet data channels or coming from other noise sources cause the O-band QKD system to fail. Noise may also occur intermittently in a real network, and in this case it might be difficult for a QKD service provider to find out the cause. We remind the reader that detection of just one noise photon out of every five detected photons on average at the receiver is enough to lead to a quantum bit-error rate (QBER) value of > 10% and prevent the QKD system from effectively producing keys. Thus noise tolerance is an important consideration when implementing O-band QKD.

It is common to employ a weak coherent source (WCS) with mean photon number $\mu < 1$ per time-slot for QKD. In the case of $\mu = 0.1$, even before transmission, 90% of the time-slots are empty slots that do not contain any useful photon. In the presence of noise photons, the receiver would have a high chance of detecting these noise photons in the empty slots and registering errors. The disadvantage with using WCS is that one has no practical means of identifying the empty slots at the transmitter.

Switching to use a heralded photon source (HPS) enables such identification since an HPS is essentially a photon-pair source. The detection of one photon of a pair at the source *heralds* the presence of the other photon of the same pair [14–19]. The system can therefore send out heralding signals that accompany only time-slots containing photons at the transmitter. The heralding signals provide the photon detectors at the receiver with trigger timings, so that the detectors stay in the off state during empty slots and as a result, detection of noise photons during these slots is avoided. However, it should be clear that empty slots due to photon loss during transmission cannot be identified by this method.

Although the benefits of using photon heralding can be understood intuitively as described above, here we discuss the prospects and limitations of this method in greater depth. We note that the use of HPS for increasing the transmission distance of decoy-state QKD has already been studied before [20–24]. In this work, we specifically discuss the noise rejecting power of photon heralding scheme and perform a proof-of-concept experiment to help in our understanding. In the experiment, we consider a scenario in which a 10-km-long optical fiber channel has been corrupted by noise photons to the extent that O-band QKD using WCS is impossible. To generate the noise experimentally, we launched an L-band continuous-wave laser together with the O-band photons. The optical power launched into the transmission fiber was 4.0 dBm, which is of the same level as some commercially available transceivers. The choice of laser wavelength does not have significance here as we use the laser only to emulate a noise source.

As we tuned the L-band laser wavelength, we observed that for certain wavelengths there were substantial amount of noise photons due to both SRS and insufficient filtering of the L-band laser photons. The presence of noise photons caused photon signal-to-noise ratio (PSNR) to drop to < 4.0, which corresponds to a QBER value of > 10.0%. Note that in a real system, there are additional errors due to implementation imperfections such as inaccurate basis alignment. If overall QBER > 11%, no key can be generated. We show that under such noise conditions, using HPS instead of WCS can improve the QBER to lower than 5.7%.

We have further incorporated wavelength-division-multiplexing (WDM) into the photon heralding scheme to increase the number of QKD photons reaching the receiver [25–28]. To the best of our knowledge, this is the first time that wavelength-multiplexed heralded photons have been generated and transmitted over optical fiber. We emphasize that the noise rejection advantage of the photon heralding scheme is not limited to O-band QKD but is also applicable to similar situations in other transmission bands.

This paper is organized as follows. Section 2 explains the concept. Section 3 describes an experiment that demonstrates the advantages of using HPS over WCS. Section 4 discusses the trade-offs involved, prospects and limitations of the photon heralding scheme. Section 5 concludes this paper. The relation between PSNR and QBER is derived in Appendix A. Appendix B describes our method to obtain the mean photon-pair number $\mu$ of an HPS from experimental measurements.

2. Concept

To simplify the discussion and to make our results more general, we do not limit ourselves to particular QKD protocol but consider only detection of photonic qubits of the correct basis choice. This is acceptable because we are interested only in comparing the use of WCS and HPS for the same QKD protocol, such as BB84, and the method can be applied to other protocols as well. In this case, it suffices to consider just the photon signal-to-noise ratio (PSNR) at the receiver for a valid time-slot, defined as

$$PSNR = \frac{P_{QKD}}{P_{noise}}, \tag{1}$$

which quantifies the quality of the photon transmission. Assuming gated photon detectors, $P_{noise}$ is the probability of detecting an incoming noise photon at the receiver in a gated time-slot, while $P_{QKD}$ is probability of detecting a QKD photon in a gated time-slot. When WCS or HPS is used as photon source, $P_{QKD}$ is determined based on the requirement to suppress multi-photon probability. It is therefore not possible to increase $P_{QKD}$ to improve PSNR at the receiver. On the other hand, $P_{noise}$ may be reduced by optical filtering but this generally leads to greater optical loss. Assuming that the noise is basis-independent and that there is no more than one noise photon per time-slot, the PSNR at the receiver is related to QBER by

$$QBER \approx \frac{1}{2(1+PSNR)}. \tag{2}$$

The derivation is given in Appendix A.

Figure 1(a) shows the conceptual model for WCS. If we use a WCS producing photons with a Poisson distribution and mean photon-number of $\mu$, the probability of detecting an arriving photon at the receiver can be expressed as

$$P_s = 1 - e^{-\alpha_{tr}\alpha_d \mu}, \tag{3}$$

where $\alpha_{tr}$ is the transmittivity of the channel and $\alpha_d$ takes into account photon detector's quantum efficiency and insertion loss of optical components at the receiver. The value of $\mu$ is normally chosen to be < 0.1. One can assume $P_s \approx \alpha_{tr}\alpha_d \mu$ since $\alpha_{tr}\alpha_d \mu \ll 1$. When noise photons are present in the channel, the QKD system sees a decrease in PSNR and a corresponding increase in the QBER. If the QBER increases beyond a certain threshold value, typically 11%, the system aborts and it is unable to generate any keys if the noisy condition persists. In this situation, switching to use HPS instead improves the PSNR at the receiver and lowers the QBER.

Figure 1(b) shows the use of an HPS, which is essentially a photon-pair source. The photon-pairs are typically produced via a spontaneous parametric scattering process such as spontaneous parametric down-conversion (SPDC) or spontaneous four-wave mixing (SFWM). By detecting one photon of each output photon-pair immediately at the source, the timing of the other photon of the same pair becomes known. The system transmits this timing information in the form of heralding signals. At the receiver, the heralding signals are used to trigger the photon detectors. In this way, the heralding signals act as temporal filter rejecting uncorrelated noise photons at the receiver. Under usual circumstances, we can assume that the output from an HPS follows a Poisson distribution with mean photon-pair number $\mu$. The probability of detecting an arriving photon at the receiver conditioned on the triggering of photon detector at the transmitter can be expressed as

$$\frac{P_c}{P_t} = \frac{1 - e^{-\beta\mu} - e^{-\alpha_s\alpha_{tr}\alpha_d \mu} + e^{(\alpha_s\alpha_{tr}\alpha_d \beta - \alpha_s\alpha_{tr}\alpha_d - \beta)\mu}}{1 - e^{-\beta\mu}}. \tag{4}$$

Here, $P_c$ is the joint probability that the photon detector is triggered by a heralding signal and at the same time it detects a photon, while $P_t$ is the probability of HPS producing a heralding signal, or *heralding probability*. The *heralding efficiency* of the HPS is defined as the probability of getting a non-empty time-slot conditioned on the triggering of photon detector at the transmitter and its expression can be obtained by setting $\alpha_{tr}\alpha_d = 1$ in Eq. (4). $\beta$ denotes the transmittivity of the idler arm which includes optical component losses and quantum efficiency of the heralding detector. The PSNR for both HPS and WCS, which we denote by $PSNR_{HPS}$ and $PSNR_{WCS}$, respectively, can be expressed as

$$PSNR_{HPS} = \frac{P_{QKD,HPS}}{P_{noise}} = \frac{P_c}{P_{noise}P_t}, \tag{5}$$

$$PSNR_{WCS} = \frac{P_{QKD,WCS}}{P_{noise}} = \frac{P_s}{P_{noise}}. \tag{6}$$

Under the same noise condition, the improvement to PSNR that could be brought about by use of photon heralding can be expressed as the ratio

$$\chi \equiv \frac{PSNR_{HPS}}{PSNR_{WCS}} = \frac{P_c}{P_t P_s} \approx \frac{\alpha_s}{\mu}(1+\mu), \tag{7}$$

where the approximation is valid if $\alpha_{tr}\alpha_d, \beta \ll 1$. For $\mu \ll 1$,

$$\chi \approx \frac{\alpha_s}{\mu}. \tag{8}$$

This result can be understood intuitively as follows. When a WCS having mean photon number $\mu$ is used, a large proportion of time-slots are empty slots that do not contain any photon. When noise photons fall into these empty time-slots, PSNR is affected. The higher the proportion of empty slots, the higher the probability of detecting a noise photon. Photon heralding reduces the proportion of empty slots if $\alpha_s > \mu$. However, in the case where $\alpha_s = \mu$, there should be little improvement to PSNR even if we replace WCS with HPS, as the proportion of empty slots remains almost the same. We also find that using HPS will lead to decrease of photon transmission rate due to both heralding probability and heralding efficiency < 1. The extent of the decrease of photon transmission rate at the transmitter can be obtained from the ratio

$$\frac{P_c}{P_s} \approx \alpha_s \beta, \tag{6}$$

where again we set $\alpha_{tr}\alpha_d = 1$. This result shows that although it is possible to increase PSNR by photon heralding, the price to pay is reduced photon rates. Therefore, it is important to improve both the heralding probability and heralding efficiency of an HPS to achieve $P_c/P_s$ as close to unity as possible. Another potential method to offset the lower photon rates is to introduce wavelength-multiplexing into the photon heralding scheme to increase overall photon rates. In our experiment, we demonstrate how this can be done.

3. Experiment

To produce the O-band wavelength-multiplexed heralded single photons, we have used a broadband photon-pair source based on a pulse-pumped silicon wire waveguide. The photon-pairs are produced in the silicon material by spontaneous four-wave mixing (SFWM). The advantages of using silicon wire waveguides over other materials, such as optical fiber, for photon-pair generation include good optical confinement, large nonlinearity, and potential for large-scale integration [29–34]. We have used a 2.6-mm-long silicon wire waveguide. The waveguide's transverse dimensions of 440 nm × 220 nm were chosen for obtaining a zero-dispersion wavelength near 1310 nm [35]. This is required for effective phase-matching of the SFWM process. Through pumping with 1310 nm laser pulses having 100 ps pulse-widths and adjustable optical powers, we obtained photon-pairs covering a broad bandwidth in the O-band. The pulse repetition rate was 48.7 MHz.

For wavelength demultiplexing, we have used a custom-made, 64-channel O-band arrayed waveguide grating (AWG) having channel spacing of 50 GHz for the signal band (< 1310 nm) and a tunable fiber Bragg grating (FBG) filter (Alnair Labs, WTF-200) for the idler band (> 1310 nm). The AWG channel wavelengths ranged from 1308.2 nm (Channel 1) to 1290.4 nm (Channel 64). The FBG filter has a tuning range of 10 nm from 1310 nm to 1320 nm.

We measured the coincidence-to-accidental ratio (CAR) of the source for seven channels and observed highest CAR value of 113.3 for Channel 11 (wavelength 1305.3 nm), as shown in Fig. 2. This shows that we can simultaneously obtain multi-wavelength O-band heralded single-photons using just one silicon wire waveguide. Waveguide propagation loss and waveguide-fiber coupling loss were measured to be 1.0 dB and 3.0 dB per facet, respectively. The low coupling loss was made possible via use of a suitable mode-size converter [36]. More details on the photon-pair source can be found in [37, 38].

Before performing the transmission experiment, we measured the second-order correlation function $g^{(2)}(0)$ of the heralded photons at wavelength 1305.3 nm in a back-to-back Hanbury-Brown–Twiss (HBT) type three-fold coincidence experiment. Figure 3 shows the setup for this experiment. We used two tunable FBG (Alnair Labs, WTF-200) filters to select the photon wavelengths. Three single-photon counter modules (SPCMs) were used. SPCM 1 (IDQ, id210) was used to detect the heralding photons. Its detection output triggers SPCMs 2 and 3 (IDQ, id201) that were used to detect the heralded photons. SPCM gate-widths were set to 2.5 ns and deadtimes were set to 10 µs. Both the 20-m-long standard single-mode fiber (SMF) placed before the 50/50 coupler and the digital delay generator (DDG, Highland Technology, P400) at SPCM 1's detection output were needed for trigger timing synchronization. The value of $g^{(2)}(0)$ was calculated from $g^{(2)}(0) = (N_t N_c)/(N_2 N_3)$, where $N_2$ and $N_3$ are the count rates of SPCMs 2 and 3, respectively. $N_t$ is the triggering rate, while $N_c$ is the coincidence count rate. Figure 4 shows the experimental result. A $g^{(2)}(0)$ value of $< 0.20$ was obtained for triggering rates $< 20$ kHz, showing that the source worked well as an HPS.

Figure 5 shows a schematic of the transmission experiment. At the transmitter, we have used one tunable FBG filter to select the heralding channel. SPCM 1 was used to detect the heralding photon and its output triggered a 1551 nm semiconductor laser (IDQ, id300) to produce heralding signals. The heralding signals were amplified using an erbium-doped fiber amplifier (EDFA, Nuphoton Technologies, NP2000) and filtered before being sent into the 10-km-long transmission fiber (Corning, SMF-28) together with the O-band photons. A wavelength-tunable L-band laser (Yenista Optics, T100 1620) was used to generate noise photons. At the receiver, an avalanche photodiode (APD, Thorlabs, APD110C) detected the filtered heralding signals and the detection output was used to gate SPCMs 2 and 3 for detecting the O-band heralded photons. The 64-channel O-band AWG was used for wavelength-demultiplexing of the heralded photons. An additional wavelength-tunable FBG filter (Alnair Labs, WTF-200) of 0.35 nm bandwidth was used to further reject out-of-band noise photons.

Figure 6 shows how noise photon count rates for Channel 16 (wavelength 1303.9 nm) depended on the L-band laser wavelength. The measurement was made before the 50/50 coupler at the receiver. Similar wavelength dependence was observed for other channels as well. This wavelength dependence was largely due to out-of-band laser photons that leaked through a free-spectral-range (FSR) mode of the AWG. Even though we cascaded an FBG filter to filter off the out-of-band photons, the suppression ratio was insufficient. In real application where dense-WDM (DWDM) channels could be added or dropped dynamically in the C- or L-band, the amount of noise photons experienced by the QKD system may be variable and not easily predictable. Cascading more optical filters will increase the optical loss experienced by the QKD photons. In our transmission experiment, we intentionally chose the L-band laser wavelengths that gave highest noise photon count rates for each measured channel for the purpose of concept demonstration.

When the L-band laser was turned off, there were no noise photons in the transmission fiber. At a heralding rate of 20 kcps, we obtained $g^{(2)}(0)$ values of 0.219, 0.228 and 0.262 for Channel 11 (wavelength 1305.3 nm), Channel 16, and Channel 21 (wavelength 1302.5 nm), respectively. When the L-band laser was turned on, the $g^{(2)}(0)$ values became 0.345, 0.253, and 0.339 for the three channels. This shows that the presence of noise photons did cause $g^{(2)}(0)$ to degrade. The greater extent of degradation for Channel 11 as compared to Channel 16, was due to higher transmission loss, whereas for Channel 21, it was due to both a lower SFWM efficiency at that wavelength, as well as slightly higher transmission loss.

The heralding rate of 20 kcps corresponds to mean photon number of $\mu = 0.11$. This is explained in more details in Appendix B. $PSNR_{WCS}$ can be estimated using Eqs. (6) and (3), where $\alpha_{tr}\alpha_d$ is calculated from Eq. (4) using known values of $\beta$, $\mu$, and $\alpha_s$. $P_{noise}$ was calculated from experimentally measured noise photon count rates. The estimated $PSNR_{WCS}$ values were 3.45, 4.06, and 3.67 for channels 11, 16, and 21, respectively. These PSNR values correspond to QBER values of 11.2%, 9.9%, and 10.7%, which were all too high for producing secure keys effectively. On the other hand, the $PSNR_{HPS}$ values were obtained from measured photon count rates and estimated $P_{noise}$ at the heralding rate. Figure 7 shows that $PSNR_{HPS}$ = 9.18 (QBER = 4.9%) for Channel 16. For Channel 11 and Channel 21, the measured $PSNR_{HPS}$ values were 7.79 (QBER = 5.7%) and 8.3 (QBER = 5.4%). This result clearly shows that using HPS for QKD leads to better noise tolerance compared to WCS. Although we did not measure the performance of all the channels between channels 11 and 21, their characteristics should not differ too much. We have thus also demonstrated the benefits of incorporating WDM into photon heralding scheme to increase the overall photon rate.

4. Discussion

It should already be obvious from Eq. (7) that although the photon heralding scheme improves the PSNR of photon transmission under noisy condition, the noise rejection power of this scheme has its limit. Even in the ideal case of $α_s$ = 1, the maximum achievable improvement to PSNR through switching from WCS to HPS is $(1+μ)/μ$. This is simply because photon heralding cannot prevent detection of noise photons that fall into non-empty time-slots and into time-slots whose photons are lost during transmission.

The heralding efficiency of our HPS was approximately 22.4%, which takes into account 3.0 dB of waveguide-fiber coupling loss, 1.0 dB of propagation loss in the silicon wire waveguide, and the optical losses of two pump-suppression FBGs and a WDM coupler. Substituting $μ$ = 0.11 into Eq. (7), we find that our HPS gives a $χ$ value of 2.26. Recently, very good performance HPS with high heralding efficiencies at 1550 nm and 810 nm have been reported [17–19]. Assuming that they can be realized at O-band, using these sources, the photon heralding scheme is expected to give even better performance. An HPS with heralding efficiency of 45% [17] would have a $χ$ value of 4.54. For a heralding efficiency of 84% [18, 19], $χ$ would be 8.48.

The improvement to PSNR that is brought about by use of HPS is not without trade-off. The photon rates are reduced due to optical losses within the HPS and also non-unity quantum efficiency of the heralding detector. We estimated $α_s$ = –6.5 dB, $β$ = –23.3 dB for the HPS that we used. According to Eq. (9), the photon rate in our experiment would suffer a reduction by –29.8 dB, which is almost 3 orders of magnitude. If the source of [17] can be used, we would have $α_sβ$ of about –13 dB. In this case, the incorporation of 20-channel-WDM into the photon heralding scheme, as shown in our experiment, would be sufficient to compensate for the drop in raw key rate. It is nevertheless still important to strive to improve both the heralding rate and heralding efficiency of HPS as much as possible.

In an actual QKD implementation, one must encode quantum information on the transmitted photons [39]. The use of optical modulator for either phase or polarization encoding introduces additional loss of 3 to 4 dB at the transmitter. If one uses WCS, one can set the mean photon number $μ$ to be 0.1 at the output of the optical modulator. For HPS, this would not be possible. Any loss due to optical modulator at the signal arm decreases the heralding efficiency. For a WDM scheme, the situation is worse. One would need to separate the wavelength channels (using an AWG, for example) for independent phase or polarization modulation before combining the wavelength channels again (using a second AWG). This incurs substantial optical loss, and heralding efficiency is reduced significantly. It is therefore more favorable to consider implementing the WDM photon heralding scheme in a wavelength-multiplexed entanglement-based QKD setting [40], where entangled photon-pairs are used. In this case, one photon of each entangled photon-pair is detected at the transmitter, while the other photon of the same pair is transmitted over the optical fiber to the receiver. There is no need to perform encoding using optical modulators at the transmitter as the correlations that exist between entangled photons in each wavelength channel can be used to generate the raw key. To incorporate photon heralding into an entanglement-based QKD scheme, one simply converts photon detection timings at the transmitter into heralding signals and sends them together with the entangled photons to the receiver. Since noise photons that reach the receiver when there is no heralding signal are not detected by the photon detectors, the proposed scheme has the additional advantage that detector dead-times associated with detection of these noise photons are avoided altogether.

5. Conclusion

Operating QKD in the O-band has the important advantage that no restrictions on Internet channel wavelength or optical power need to be imposed. However, noise photons due to SRS coming from Internet data channels propagating in the same fiber or other unexpected noise sources, such as photons that leak through an FSR mode of AWG, may lead to corruption of the QKD channel. It is therefore crucial to improve the noise tolerance of QKD. We have shown the benefits of using an HPS instead of a WCS in reducing the noise photon detection probability at the receiver. The feasibility of this idea was demonstrated in a proof-of-concept experiment, in which we successfully transmitted WDM O-band heralded photons over 10 km of optical fiber under noisy condition. The measured PSNR values at the receiver were all > 7.8 which corresponds to a QBER of < 5.7%. This is good enough for distilling secure keys. On the other hand, if we had used a WCS under the same noise condition, the QBER would be too high for secure key generation. Our calculations suggest that the use of recently developed HPS with higher heralding rates and heralding efficiencies [17–19] compared to ours could lead to even better performance. The improved noise tolerance of O-band QKD via use of the photon heralding scheme is a significant step towards widespread non-disruptive deployment of QKD systems in existing fiber-optic networks.

Appendix A: Relation between PSNR and QBER

Let us define photon signal-to-noise ratio (PSNR) as the ratio between the number of arriving QKD photons and the number of arriving noise photons at the QKD receiver. It can be expressed in terms of probability of detecting a QKD photon, $P_{QKD}$, and probability of detecting a noise photon, $P_{noise}$, as

$$PSNR = \frac{P_{QKD}}{P_{noise}}. \quad (10)$$

In any QKD scheme that is based on photonic qubits, two photon detectors at the receiver are used to obtain one bit value for each detected photon in the correct basis. The probability of registering an error and the probability of registering no error are given by

$$P_{error} = (1 - P_{QKD})\frac{P_{noise}}{2} + P_{QKD}\frac{P_{noise}}{2}\frac{1}{2}, \quad (11)$$

$$P_{no\,error} = (1 - P_{QKD})\frac{P_{noise}}{2} + P_{QKD}\left(1 - \frac{P_{noise}}{4}\right), \quad (12)$$

respectively. Here we assume that the noise is basis-independent and that there is no more than one arriving noise photon per time-slot. The second assumption becomes invalid when $P_{noise}$ becomes larger than $P_{QKD}$, but in this case QBER > 11% and so we are not interested. The factor of 1/2 in the second term of Eq. (11) accounts for random bit value assignment when both detectors click simultaneously. The QBER is therefore related to PSNR by the following expression

$$QBER = \frac{P_{error}}{P_{error} + P_{no\,error}} \approx \frac{1}{2(1 + PSNR)}, \quad (13)$$

where we have used the fact that $P_{QKD}$ is almost always $\ll 1$. It is obvious from Eq. (13) that a PSNR of > 9 is needed to obtain QBER of lower than 5%.

Appendix B: Method to obtain $\beta$ and $\mu$ from experimentally observed $g^{(2)}(0)$ and heralding rate $r$

Our single-photon counting modules (SPCMs) have a deadtime of 10 μs. Correcting for the effect of SPCM deadtimes, which we denote by $\tau_d$, we can write down the expression

$$\beta\mu = \frac{r}{(1-\tau_d r)F}, \quad (14)$$

where $r$ is the experimentally observed heralding rate and $F$ is the repetition rate of the pump pulses. The second-order correlation function $g^{(2)}(0)$ is approximately [17, 41]

$$g^{(2)}(0) = \frac{2\mu - \beta\mu + \mu^2}{1 + 2\mu - \beta\mu + \mu^2}. \quad (15)$$

We can therefore solve for $\mu$ using experimental values of $g^{(2)}(0)$. In this way, we have obtained $\beta = -23.3$ dB, and for a heralding rate of $r = 20$ kcps, mean photon-pair number $\mu = 0.11$.

Figures and Captions

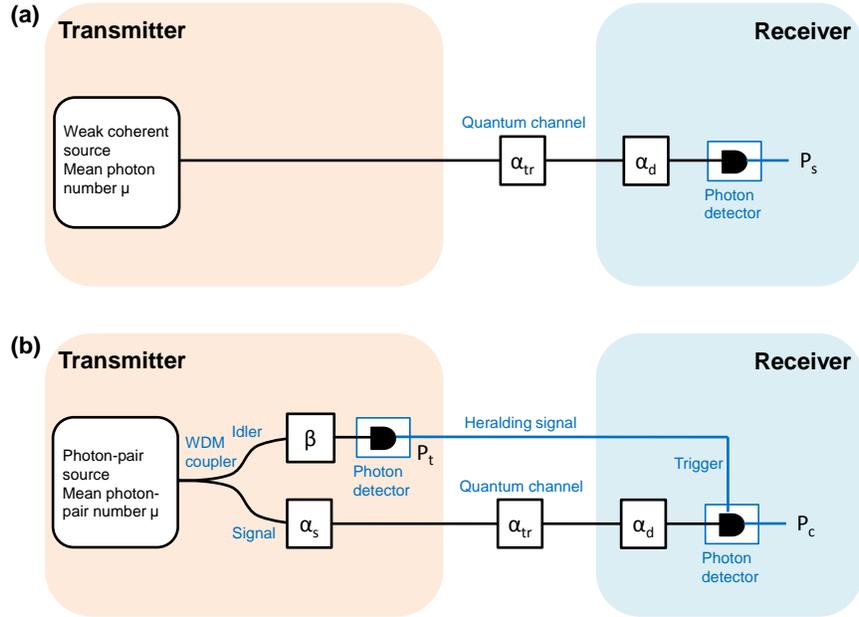

Fig. 1. Conceptual models for theoretical calculation. Channel transmittivity is denoted by $\alpha_{tr}$ and photon detector efficiency is denoted by $\alpha_d$. Effects of photon detector imperfections such as dark counts, afterpulsing, and deadtimes are omitted in the models for simplicity. (a) For weak coherent source (WCS), the probability of detecting a photon at the receiver is denoted by $P_s$. (b) For a heralded photon source (HPS) having mean photon-pair number of $\mu$, the probability of producing a heralding signal at the transmitter is denoted by $P_t$. At the receiver, $P_c$ denotes the joint probability of arrival of a heralding signal and detection of a photon by the photon detector. The signal and idler arms of the source have transmittivities of $\alpha_s$ and $\beta$, respectively.

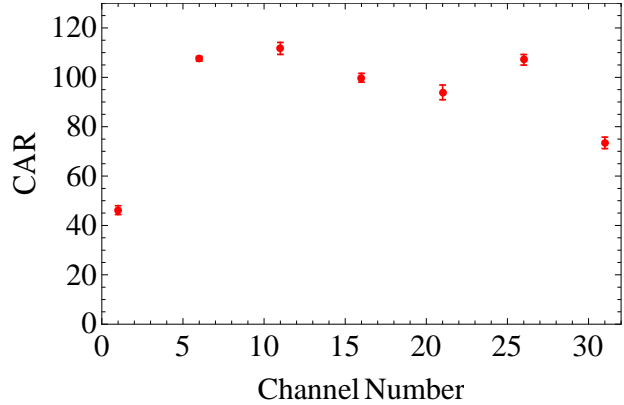

Fig. 2. Measured coincidence-to-accidental ratio (CAR) for AWG channels from Channel 1 (wavelength 1308.2 nm) to Channel 31 (wavelength 1299.6 nm) with a measurement spacing of 5 channels. The pump power was chosen to allow observation of highest CAR value at Channel 11 (wavelength 1305.3 nm). The highest observed CAR value was 113.3 for this channel. A lower CAR value for Channel 1 was due to leakage of pump photons, while for Channel 31, it was due to Raman scattering photons generated inside the silicon waveguide, and they leaked through a free-spectral range (FSR) mode of the AWG [37].

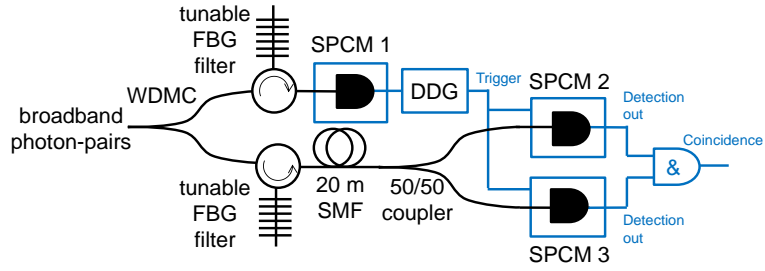

Fig. 3. Back-to-back measurement of second-order correlation function using a Hanbury-Brown–Twiss (HBT) setup. FBG: fiber Bragg grating; WDMC: wavelength-division-multiplexing coupler; SPCM: single-photon counter module.

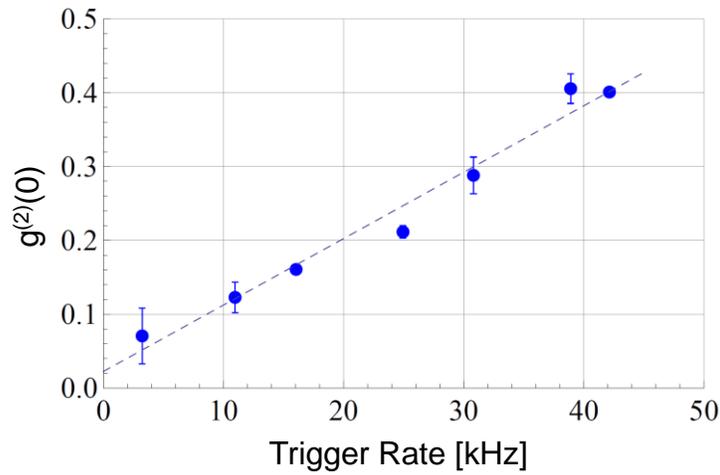

Fig. 4. Measured $g^{(2)}(0)$ values at heralded photon wavelength of 1305.3 nm. At triggering rates < 20 kHz, we obtained $g^{(2)}(0)$ values of < 0.2. Dashed line is linear fit.

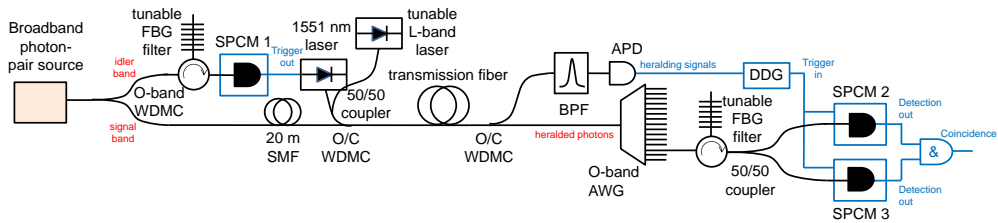

Fig. 5. Schematic of the transmission experiment. The receiver implements a Hanbury-Brown-Twiss (HBT) type measurement to measure the second-order correlation function. APD: avalanche photodiode; AWG: arrayed waveguide grating; BPF: band-pass filter; DDG: digital delay generator; FBG: fiber Bragg grating; O/C: O-band/C-band; SMF: single-mode fiber; SPCM: single-photon counter module; WDMC: wavelength-division-multiplexing coupler.

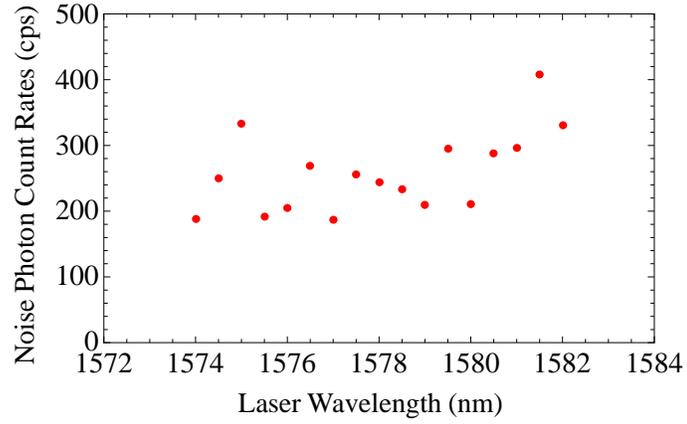

Fig. 6. Measured noise photon count rates for Channel 16 (wavelength 1303.9 nm) versus L-band laser wavelength. Laser power launched into transmission fiber was 4.0 dBm. Detector clock rate was 1 MHz.

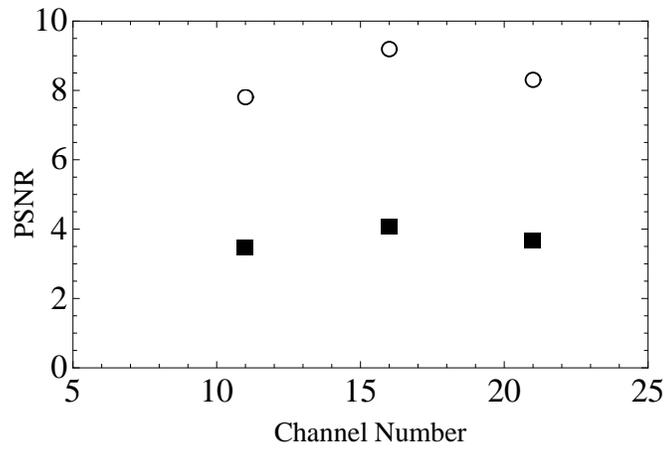

Fig. 7. $PSNR_{HPS}$ (open circles) and $PSNR_{WCS}$ (filled squares) for three selected wavelength channels, 11, 16, and 21.